\documentclass[12pt,a4paper]{article}
\usepackage{mathrsfs}
\usepackage{epsfig}
\usepackage{slashed}
\begin{document}

\title{Precise photoproduction of the charged top-pions at the LHC with forward detector acceptances}
\author{Hao Sun$^{1,3}$\footnote{haosun@mail.ustc.edu.cn \hspace{0.2cm} haosun@dlut.edu.cn},
        Chong-Xing Yue$^2$\footnote{cxyue@lnnu.edu.cn} \\
{\small $^{1}$ Department of Physics, Dalian University of Technology, Dalian 116024, P.R.China} \\
{\small $^{2}$ Department of Physics, Liaoning Normal University, Dalian 116029, P.R.China}\\
{\small $^{3}$ School of Physics and Technology, University of Jinan, Jinan 250022, P.R.China} \\
}
\date{\today}
\maketitle

\begin{abstract}

We study the photoproduction of the charged top-pion
predicted by the top triangle moose ($TTM$) model
(a deconstructed version of the topcolor-assisted technicolor $TC2$ model)
via the processes $pp\rightarrow p \gamma p \rightarrow \pi^\pm_t t +X$
at the 14 $TeV$ Large Hadron Collider ($LHC$)
including next-to-leading order ($NLO$) $QCD$ corrections.
Our results show that the production cross sections and distributions are sensitive
to the free parameters $\sin\omega$ and $M_{\pi_t}$.
Typical $QCD$ correction value is $7\% \sim 11\%$ and does not depend much on $\sin\omega$ as well as
the forward detector acceptances.

\vspace{0.5cm} \vspace{2.0cm} \noindent
\end{abstract}

\newpage
\section{Introduction}
The top quark is the heaviest known elementary particle which
makes it an excellent candidate for new physics searches.
Origin of its mass might be different from that of other quarks and leptons,
a top quark condensate ($<t\bar{t}>$), for example,
could be responsible for at least part of the mechanism of electroweak symmetry breaking ($EWSB$).
An interesting model involving a role for the top quark in dynamical $EWSB$
is known as the topcolor-assisted technicolor ($TC2$) model\cite{TC2}.
Higgsless models\cite{higgsless} have emerged as a novel way of understanding the mechanism
of $EWSB$ without the presence of a scalar particle in the spectrum.
Recently, combing Higgsless and topcolor mechanisms,
a deconstructed Higgsless model was proposed,
called the top triangle moose ($TTM$) model\cite{TTM0,TTM1}.
In this model, $EWSB$ results largely from the Higgsless mechanism
while the top quark mass is mainly generated by the topcolor mechanism.
The $TTM$ model alleviates the tension between obtaining the correct top quark mass
and keeping $\Delta\rho$ small that exists in many Higgsless models, which can be
seen as the deconstructed version of the $TC2$ model.
The new physics models belonging to the topcolor scenario genetically have two sources
of $EWSB$ and there are two sets of Goldstone bosons.
One set is eaten by the electroweak ($EW$)
gauge bosons $W$ and $Z$ to generate their masses, while the other set remains in the
spectrum, which is called the top-pions ($\pi^0_t$ and $\pi^\pm_t$).
Topcolor scenario also predicts the existence of the top-Higgs $h^0_t$ ,
which is the $t\bar{t}$ bound state.
The possible signals of these new scalar particles have been extensively studied
in the literature, however, most are done in the context of the $TC2$ model.
Phenomenology analysis about the top-pions and top-Higgs predicted
by the $TTM$ model\cite{TTM0,TTM1,TTM} is necessary.

The Large Hadron Collider ($LHC$) generates high energetic proton-proton ($pp$) collisions
with a luminosity of ${\cal L}=10^{34}cm^{-2}s^{-1}$. It provides high statistics data at high energies.
On the other hand hadronic interactions generally involve serious backgrounds which should be concerned.
A new phenomenon called exclusive production was observed in the
measurements of $CDF$ collaboration include
the exclusive lepton pairs production\cite{ppllpp}, photon photon production\cite{pprrpp},
dijet production\cite{ppjjpp}, the exclusive charmonium ($J/\psi$) meson photoproduction\cite{ppJPHIpp}, etc.
Complementary to $pp$ interactions, studies of exclusive production of leptons, photon
and heavy particles might be possible and opens new field of
studying very high energy photon-photon ($\gamma\gamma$) and photon-proton ($\gamma p$) interactions.

Following the experience from $HERA$ and the $Tevatron$ new detectors
are proposed to be installed in the $LHC$ tunnel
as an additional upgrade of the $ATLAS$ and $CMS$ detectors.
They have a program of forward physics with extra detectors
located in a region nearly 100m-400m from the interaction point.
These forward detector equipment allows one to detect intact scattered protons after the collision.
Therefore the processes which spoil the proton structure,
can be easily discerned from the exclusive photo-production processes.
By use of forward detector equipment we can eliminate many serious backgrounds.
This is one of the advantages of the exclusive photo-production processes.

A brief review of experimental prospects for studying high-energy
$\gamma\gamma$ and $\gamma p$ interactions are discussed in Ref.\cite{HEPhotonIntatLHC}
and cross sections are calculated for many $EW$ and $BSM$ processes.
Many other phenomenological studies on photoproduction processes involve:
supersymmetry\cite{SUSYprrp1,SUSYprrp2}, extra dimensions\cite{EDprrp,EDpllp},
unparticle physics\cite{unparticle}, gauge boson self-interactions
\cite{AnomalousZZrr,AnomalousWWr1,AnomalousWWr2,AnomalousWWrr, AnomalousZZZ},
neutrino electromagnetic properties\cite{electromagnetic1,electromagnetic2,electromagnetic3},
the top quark physics\cite{Anomaloustqr,AnomalousWtb}, etc.

Photoproduction of the charged top-pion at leading order ($LO$) has been studied in Ref.\cite{topionrcpt}
which proceeds via the subprocess $\gamma c\rightarrow \pi^\pm_t b$ mediated by the flavor
changing couplings and  through $\gamma b \rightarrow \pi^\pm_t t$
at the large hadron-electron collider ($LHeC$)\cite{topionrbpt}.
At the $LHC$, in general $pp$ collision, the charged top-pion can be produced
in association with a top quark through bottom-gluon fusion,
$gb\rightarrow t \pi^-_t$ , and through gluon-gluon fusion, $gg\rightarrow \bar{b}t\pi^-_t$,
phenomenologically similar to a charged Higgs boson in a two-Higgs-doublet model with low tan$\beta$.
Related $NLO$ study can be find in Ref.\cite{chargedPiQCDLHC}.
On the other hand, $\pi^\pm_t$ t associated production at the $\gamma p$ collision $LHC$ will
be very clean or at least with backgrounds easy going, thus leading a good chance to be detected.
It can be a complementary process to be studied in addition of $gb\rightarrow t \pi^-_t$.
In this paper, we present this production at the $\gamma p$ collision
assuming a typical $LHC$ multipurpose forward detector.
Accurate theoretical predictions including higher order $QCD$ corrections are included.
Paper is organized as follows: in section 2 we present a brief
introduction to the calculation framework including the $TTM$ model description,
$EPA$ implementation and $LO$ and $NLO$ cross section calculations.
Section 3 is arranged to present the numerical checks and results of our studies.
Finally we summarize the conclusions in the last section.

\section{Calculation Framework}

\subsection{The essential features of the $TTM$ model}

The detailed description of the $TTM$ model can be found in Refs.\cite{TTM0,TTM1},
and here we just briefly review its essential features, which are related to our calculation.
The $EW$ gauge structure of the $TTM$ model is
$SU(2)_{0}\times SU(2)_{1}\times U(1)_{2}$. The nonlinear sigma
field $\sum_{01}$ breaks the group $SU(2)_{0}\times SU(2)_{1}$ down to
$SU(2)$ and field $\sum_{12}$ breaks $SU(2)_{1}\times U(1)_{2}$
down to $U(1)$. To separate top quark mass generation from $EWSB$, a
top-Higgs field $\Phi$ is introduced to the $TTM$ model, which
couples preferentially to the top quark. To ensure that most of  $EWSB$
comes from the Higgsless side, the $VEVs$ of the fields $\sum_{01}$
and $\sum_{12}$ are chosen to be $<\sum_{01}>=<\sum_{12}>=F=\sqrt2
\nu\cos\omega$, in which $\nu=246 GeV$ is the $EW$ scale and
$\omega$ is a new small parameter. The $VEV$ of the top-Higgs field
is $f=<\Phi>=\nu \sin\omega$.

From above discussions, we can see that, for the $TTM$ model, there
are six scalar degrees of freedom on the Higgsless sector and four
on the top-Higgs sector. Six of these Goldstone bosons are eaten to
give masses to the gauge bosons $W^{\pm}$, $Z$, $W'^{\pm}$ and $Z'$.
Others remain as physical states in the spectrum, which are called
the top-pions ($\pi_{t}^{\pm}$ and $\pi_{t}^{0}$) and the top-Higgs
$h_{t}^{0}$. In this paper, we will focus our attention on photoproduction
of the charged top-pions via $\gamma p$ collisions at the  $LHC$. The
couplings of the charged top-pions $\pi_{t}^{\pm}$ to ordinary
particles, which are related to our calculation, are given by Ref.\cite{TTM1}

\begin{eqnarray}
{\cal L}_{\pi_{t}tb}=i\lambda_{t}\cos\omega\{1-\frac{x^{2}[a^{4}+(a^{4}-2a^{2}+2)\cos2\omega]}{8(a^{2}-1)^{2}}\}
 \pi_{t}^{+} \bar{t}_R b_{L}  + h.c.
\end{eqnarray}
with
\begin{eqnarray}
\lambda_{t}=\frac{\sqrt{2}m_{t}}{\nu\sin\omega}[\frac{M_{D}^{2}(\varepsilon_{L}^{2}+1)
-m_{t}^{2}}{M_{D}^{2}-m_{t}^{2}}],
\hspace{0.5cm} a=\frac{\nu\sin\omega}{\sqrt{2}M_{D}},\hspace{0.5cm}
x=\sqrt{2}\varepsilon_{L}=\frac{2\cos\omega M_{W}}{M_{W'}}.
\end{eqnarray}
Here we assume the $CKM$ matrix to be identity and omit the light quark masses.
$M_{D}$ is the mass scale of
the heavy fermion and $M_{W'}$ is the mass of the new gauge boson
$W'$. Since the top quark mass depends very little on the
right-handed delocalization parameter $\varepsilon_{tR}$, we have
set $\varepsilon_{tR}=0$ in $Eq.(1)$. The parameter
$\varepsilon_{L}$ describes the degree of delocalization of the
left-handed fermions and is flavor universal, the parameter $x$
presents the ratio of gauge couplings. The relationship between
$\varepsilon_{L}$ and $x$, which is given in  $Eq.(2)$, is imposed by ideal delocalization.

Ref.\cite{TTM2} has shown that $M_{W'}$ should be larger than 380 $GeV$
demanded by the $LEPII$ data and smaller than 1.2 $TeV$ by the
need to maintain  perturbative unitarity in $W_{L} W_{L}$ scattering.
It is obvious that the coupling $\pi_{t}tb$ is not very sensitive to
the parameters $M_{W'}$ and $M_{D}$. Thus, the production cross sections of the subprocesses
$\gamma b\rightarrow t \pi_{t}^{-}$ and $\gamma \overline{b}\rightarrow \overline{t}
\pi_{t}^{+}$ are not strongly depend on the values of the mass parameters $M_{W'}$ and $M_{D}$.
In our following numerical
calculation, we will take the illustrative values $M_{W'}=500GeV$
and $M_{D}=400GeV$. In this case, there is
$[M_{D}^{2}(\varepsilon_{L}^{2}+1)-m_{t}^{2}]/(M_{D}^{2}-m_{t}^{2})\approx1$
and Eq.(1) can be approximately written as
\begin{eqnarray}
{\cal L}_{\pi_{t}tb}\approx i\frac{\sqrt{2}m_{t}C}{\nu}\cot\omega
 \pi_{t}^{+}\bar{t}_R b_L  + h.c.
\end{eqnarray}
with
\begin{eqnarray}
C=1-\frac{x^{2}[a^{4}+(a^{4}-2a^{2}+2)\cos2\omega]}{8(a^{2}-1)^{2}}.
\end{eqnarray}
It is obvious that constant $C$ is not sensitive to the value of $\sin\omega$ and its value close to 1.
The parameter $\sin\omega$ indicates the fraction of $EWSB$ provided
by the top condensate. The top-pion mass $M_{\pi_t}$ depends on the amount of top-quark mass
arising from the the extended technicolor ($ETC$) sector and on the effects of $EW$ gauge interactions\cite{EGI},
and thus its value is model-dependent. In the context of the $TTM$ model, Ref.\cite{TTM1} has
obtained the constraints on the top-pion mass via studying its effects on the relevant
experimental observables. Similarly with Refs.\cite{TTM1, TTM4}, we will assume it as a
free parameter.

\subsection{Equivalent Photon Approximation ($EPA$)}

In $\gamma p$ collisions, the quasi-real photons are emitted from protons with very low virtuality
so that it's a good approximation to assume that they are on-mass-shell.
These quasi-real photons scattered with small angles and low transverse momentum.
At the same time, protons emitting photons remain intact and are not spoilt.
Intact protons thus deviate slightly from their trajectory along the beam path
without being detected by central detectors.
Deflected protons and their energy loss will be detected by the forward detectors
with a very large pseudorapidity. Photons emitted with small angles by the protons
show a spectrum of virtuality $Q^2$ and the energy $E_\gamma$.
This is described by the equivalent photon approximation ($EPA$)\cite{EPA}
which differs from the point-like electron (positron) case
by taking care of the electromagnetic form factors in the equivalent
$\gamma$ spectrum and effective $\gamma$ luminosity:
\begin{equation}
 \frac{dN_\gamma}{dE_\gamma dQ^2}=\frac{\alpha}{\pi}\frac{1}{E_\gamma Q^2}[(1-\frac{E_\gamma}{E})(1-\frac{Q^2_{min}}{Q^2})F_E
 + \frac{E^2_\gamma}{2 E^2}F_M]
\end{equation}
with
\begin{eqnarray} \nonumber
 Q^2_{min}=\frac{M^2_p E^2_\gamma}{E(E-E_\gamma)}, ~~~~ F_E= \frac{4 M^2_p G^2_E + Q^2 G^2_M}{4 M^2_p +Q^2}, \\\nonumber
 G^2_E=\frac{G^2_M}{\mu^2_p}=(1+\frac{Q^2}{Q^2_0})^{-4}, ~~~~F_M=G^2_M, ~~~~Q^2_0=0.71 GeV^2 ,
\end{eqnarray}
where $\alpha$ is the fine-structure constant, E is the energy of the incoming proton beam
which is related to the quasi-real photon energy by $E_\gamma=\xi E$ and $M_p$ is the mass of
the proton. $\xi = (|p| - |p'|)/|p| $, where $p$ and $p'$ are momentums of incoming protons
and intact scattered protons, respectively. $\mu^2_p$ = 7.78 is the magnetic moment of the proton. $F_E$ and $F_M$
are functions of the electric and magnetic form factors.
In this case, if both incoming emitted protons remain intact provides the $\gamma\gamma$ collision
and it can be cleaner than the $\gamma p$ collision,
however, $\gamma p$ collisions have higher energy and effective luminosity
with respect to $\gamma\gamma$ interactions.

\subsection{The cross sections up to NLO }

We denote the parton level process as $\gamma(p_1) b(p_2)\rightarrow \pi^\pm_t(p_3) t(p_4)$
where $p_i$ are the particle four momentums.
The hadronic cross section at the $LHC$ can be converted
by integrating $\gamma b \rightarrow \pi^\pm_t t$ over the
photon($dN(x,Q^2)$) and quark($G_{b/p}(x_2,\mu_f)$) spectra:
\begin{equation}
\sigma=\int^{\sqrt{\xi_{max}}}_{\frac{M_{inv}}{\sqrt{s}}} 2z dz \int^{\xi_{max}}_{Max(z^2,\xi_{min})}
\frac{dx_1}{x_1} \int^{Q^2_{max}}_{Q^2_{min}} \frac{dN_\gamma(x_1)}{dx_1dQ^2} G_{b/p}(\frac{z^2}{x_1}, \mu_f)
\cdot \int \frac{1}{\textit{avgfac}} \frac{|{\cal M}_n ( \hat s =z^2 s
)|^2}{2 \hat s (2 \pi)^{3n-4}} d\Phi_n ,
\end{equation}
where $x_1$ is the ratio between scattered quasi-real photons and incoming proton energy
$x_1 = E_\gamma/E$ and $x_2$ is the momentum fraction of the protons momentum carried by the
bottom quark. The quantity $\hat s = z^2 s$ is the effective $c.m.s.$ energy with $z^2=x_1 x_2$.
$M_{inv}$ is the total mass of the $\pi^\pm_t t$ final state.
$\frac{2z}{x_1}$ is the Jacobian determinant when transform the differentials from $dx_1dx_2$ into $dx_1dz$.
$G_{b/p}(x,\mu_f)$ represent the bottom quark
parton density functions, $\mu_f$
is the factorization scale which can be chosen equal the
renormalization scale $\mu_r$ when the loop calculation is included.
$\frac{1}{\textit{avgfac}}$ is the times of spin-average factor,
color-average factor and identical particle factor. $|{\cal M}_n|^2$
presents the squared n-particle matrix element and divided by the
flux factor $[2 \hat s (2 \pi)^{3n-4}]$. $d\Phi_n$ is the
n-body phase space differential.

\begin{figure}[hbtp]
\vspace{-3cm}
\centering
\includegraphics[scale=0.6]{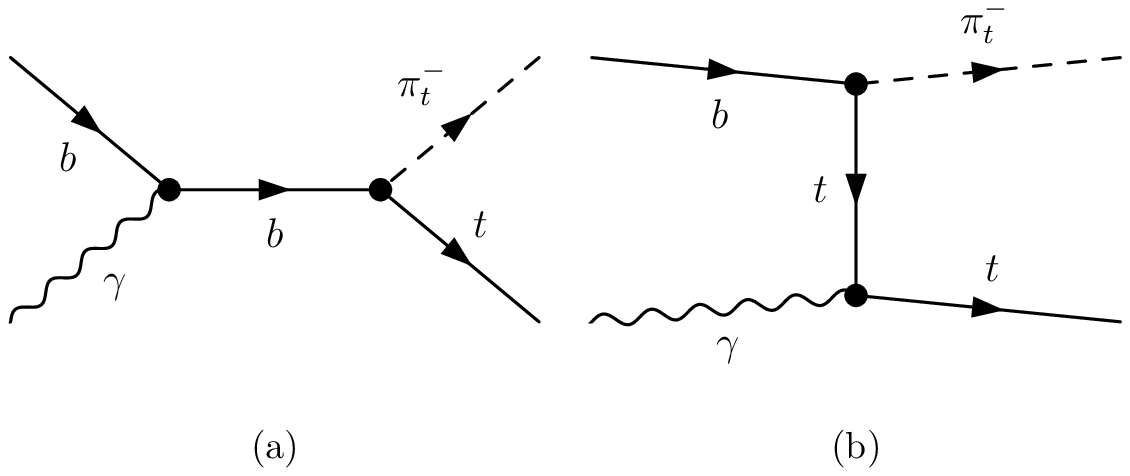}
\vspace{-12cm}
\caption{\label{fig1} Tree parton level Feynman diagrams for
$rb\rightarrow \pi^-_t t$ in the $TTM$ frame.}
\end{figure}

\begin{figure}[hbtp]
\vspace{-3.5cm}
\centering
\includegraphics[scale=0.6]{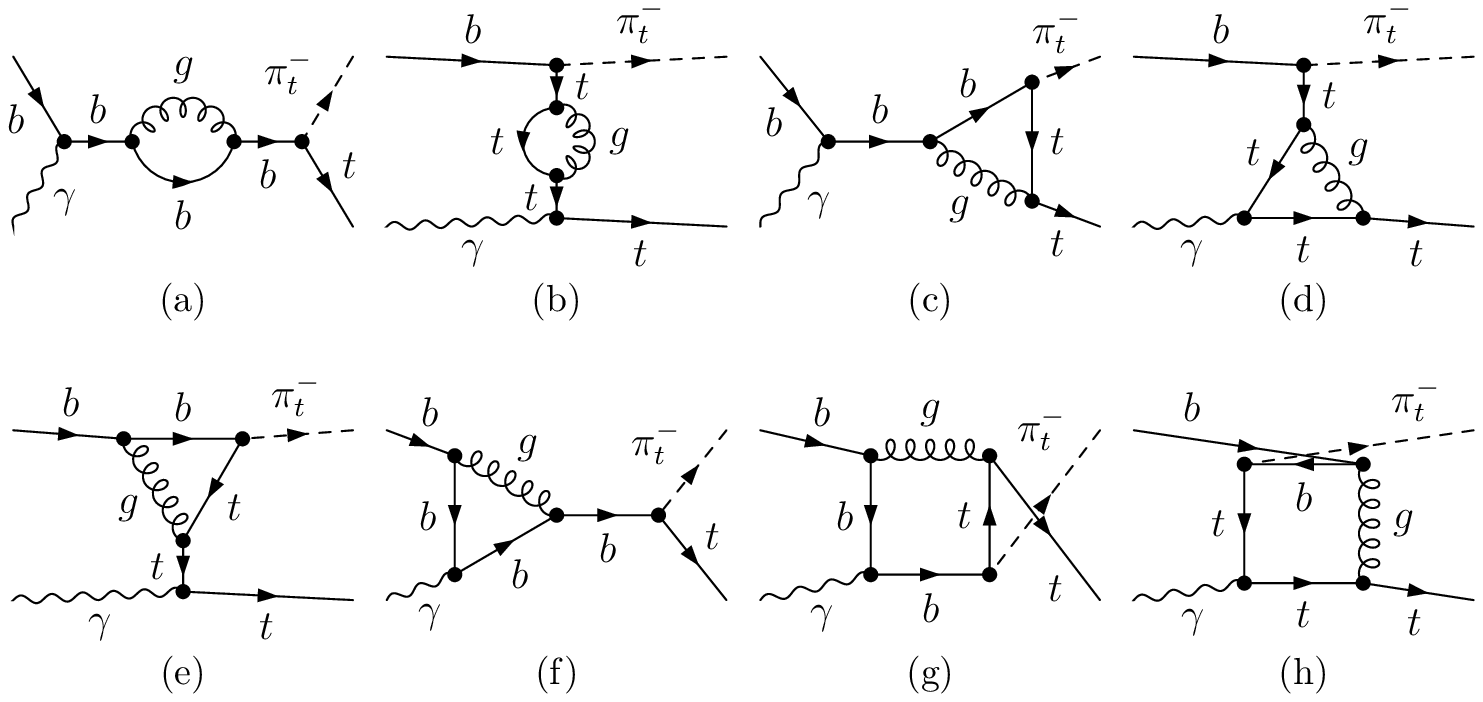}
\vspace{-9.5cm}
\caption{\label{fig2} The QCD one-loop Feynman
diagrams for the partonic process $\gamma b \rightarrow \pi^-_t t$(a-h). Counterterm diagrams correspond to
Fig.\ref{fig1} are not shown here.}
\end{figure}

\begin{figure}[hbtp]
\vspace{-3.5cm}
\centering
\includegraphics[scale=0.6]{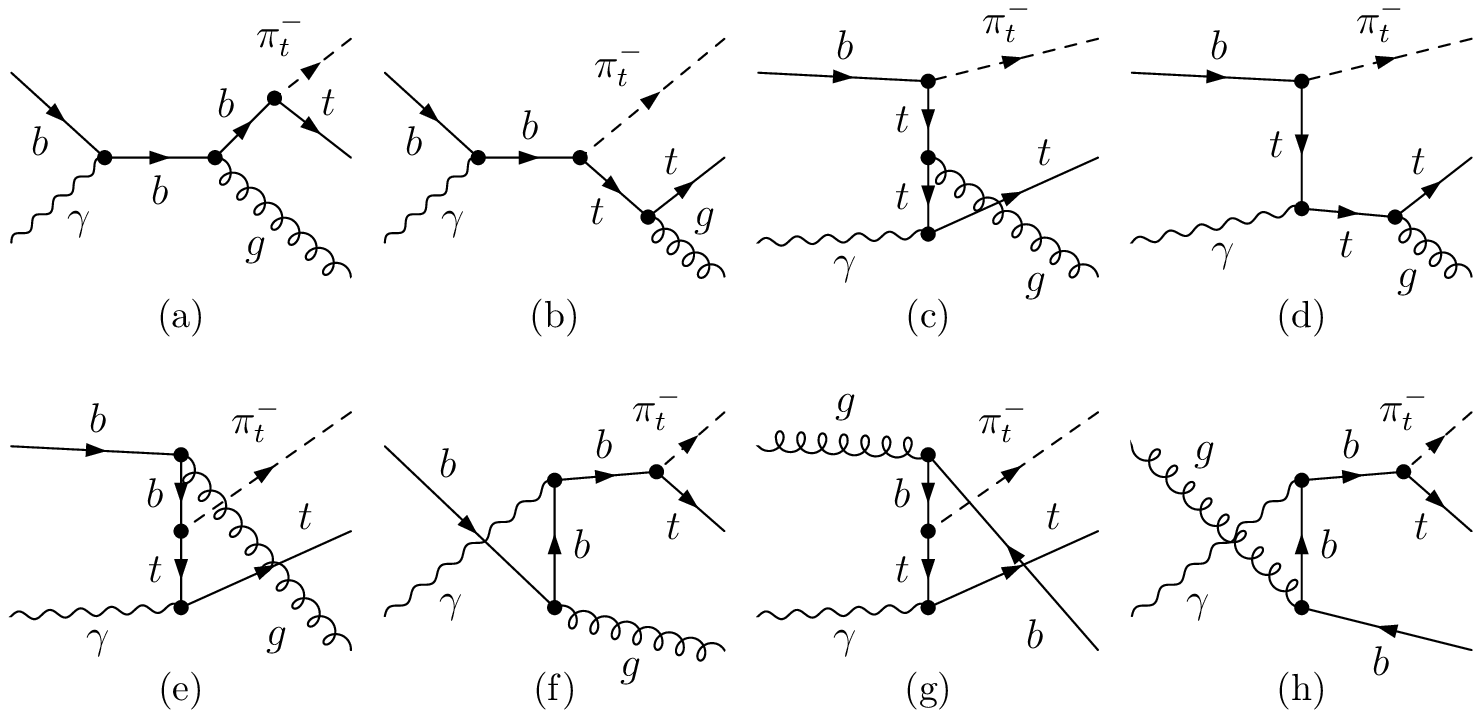}
\vspace{-9.5cm}
\caption{\label{fig3}
The tree level Feynman diagrams for
the real gluon/light-(anti)quark emission subprocess
$\gamma b \rightarrow \pi^-_t t g$ related to the first process in
Eq.\ref{Eq.hard}(a-f) and
$\gamma g\rightarrow \pi^-_t t \bar{b}$ related to the second process in
Eq.\ref{Eq.hard}(g,h).}
\end{figure}

The parton Feynman diagrams at tree level are shown in Fig.1(a,b).
We only consider the $\pi^-_t t$ production while its charge-conjugate contribution is the same.
At $NLO$ $QCD$ loop level, the Feynman diagrams are presented in Fig.\ref{fig2} and Fig.\ref{fig3},
correspond to loop ($\sigma^{loop}$) and real ($\sigma^{real}$) contributions, respectively.
There exist ultraviolet ($UV$) and soft/collinear $IR$ singularities in $\sigma^{loop}$.
To remove the $UV$ divergences, we introduce the wave function renormalization constants
$\delta Z_{\psi_{q,L,R}}$ for massless bottom and massive top fields as
$\psi^{0}_{q,L,R}=(1+\delta Z_{\phi_{q,L,R}})^{\frac{1}{2}} \psi_{q,L,R}$.
In the modified minimal subtraction ($\overline{MS}$) renormalization scheme the renormalization constants for the
massless quarks, and massive top quark (defined on shell) are expressed as
$\delta Z_{\psi_{q,L}} = -\frac{\alpha_{s}}{4\pi}C_{F}(\Delta_{UV}-\Delta_{IR})$,
$\delta Z_{\psi_{q,R}} = -\frac{\alpha_{s}}{4\pi}C_{F}\left(\Delta_{UV}-\Delta_{IR}\right)$ and
$\frac{\delta m_t}{m_t} = -\frac{\alpha_s}{3\pi} [3 \Delta_{UV} + 4]$, with $C_{F}=\frac{4}{3}$. $\Delta_{UV,IR}=\frac{1}{\epsilon_{UV,IR}}\Gamma(1+\epsilon_{UV,IR})(4\pi)^{\epsilon_{UV,IR}}$
refer to the $UV$ and $IR$ divergences, respectively. By adding renormalization part to the virtual corrections, any
$UV$ singularities are regulated leaving soft/collinear $IR$ singularities untouched.
These $IR$ singularities will be removed by combine the real emission corrections.
Singularities associated with initial state collinear gluon emission
are absorbed into the definition of the parton distribution functions.
We employ the $\overline{MS}$ scheme for the parton
distribution functions. Similar to the virtual part, we utilize
dimensional regularization to control the singularities of the
radiative corrections, which are organized using the two cutoff
phase space slicing ($TCPSS$) method\cite{2PSS:Owens}. We adopt $TCPSS$
to isolate the $IR$ singularities by introducing two cutoff parameters
$\delta_{s}$ and $\delta_{c}$. An arbitrary small $\delta_{s}$
separates the three-body final state phase space into two regions:
the soft region ($E_{5}\leq \delta_{s}\sqrt{\hat{s}}/2$) and the
hard region ($E_{5}>\delta_{s}\sqrt{\hat{s}}/2$). The $\delta_{c}$
separates hard region into the hard collinear ($HC$) region and hard
noncollinear ($\overline{HC}$) region. The criterion for separating
the $HC$ region is described as follows: the region for real
gluon/light-(anti)quark emission with $\hat{s}_{15}$ (or
$\hat{s}_{25}$) $< \delta_{c}\hat{s}$ (where
$\hat{s}_{ij}=(p_{i}+p_{j})^{2}$) is called the $HC$ region.
Otherwise it is called the $\overline{HC}$ region where in our case
related to
\begin{eqnarray}\nonumber
&&\ \ \gamma(p_1) b(p_2) \rightarrow \pi^-_t(p_3) t(p_4) g(p_5) \\
&&\ \ \gamma(p_1) g(p_2) \rightarrow \pi^-_t(p_3) t(p_4) \bar{b}(p_5)
\label{Eq.hard}
\end{eqnarray}
correspond to real gluon emission and real light-(anti)quark emission partonic processes, respectively.
After combining all these contributions above, the $UV$ and $IR$
singularities in $\sigma^{total}=\sigma^{born}+\sigma^{loop}+\sigma^S+\sigma^{HC}+\sigma^{\overline{HC}}$
are exactly canceled. Dependence on
the arbitrary small cutoff parameters $\delta_{s}$ and $\delta_{c}$
are then vanished. These cancelations can be verified numerically
in our numerical calculations.

\section{Numerical Results and Discussions}

We use FeynArts, FormCalc and our modified LoopTools (FFL)\cite{FeynArts,FormCalc,LoopTools}
packages to perform the numerical calculation.
We use CT10\cite{CT10} for the parton distributions for collider physics
and BASES\cite{BASES} to do the phase space integration.
In the numerical calculations, we take the
input as $M_p=0.938272046~{\rm GeV}$, $M_Z=91.1876~{\rm GeV}$,
$M_W=80.385~{\rm GeV}$, $M_t=173.5~{\rm GeV}$,
$\alpha(M_Z^2)^{-1}=127.918$\cite{2012PDG}, $\sqrt{s}=14~{\rm TeV}$.
For the strong coupling constant $\alpha_s(\mu)$,
we use the two-loop evolution of it with the $QCD$
parameter $\Lambda^{n_f=5}$ = 226 $MeV$ and get $\alpha_s(\mu_0)$ =
0.113. $N_f$ is the number of the active flavors.
We choose two sets of the parameters related to $TTM$ model:
\begin{itemize}
 \item Scenario 1: $M_D=400\ GeV$, $\sin\omega=0.5$, $M_{W'}=500\ GeV$, $M_{\pi_t}=400\ GeV$
 \item Scenario 2: $M_D=400\ GeV$, $\sin\omega=0.2$, $M_{W'}=500\ GeV$, $M_{\pi_t}=200\ GeV$
\end{itemize}
correspond to high (low) $M_{\pi_t}$ regions, respectively.
The detected acceptances are chosen to be\cite{AnomalousWWr1,AnomalousWWr2,xi123}:
\begin{itemize}
 \item $\xi_1$: $CMS-TOTEM$ forward detectors with $0.0015<\xi<0.5$
 \item $\xi_2$: $CMS-TOTEM$ forward detectors with $0.1<\xi<0.5$
 \item $\xi_3$: $AFP-CMS$ forward detectors with $0.0015<\xi<0.15$.
\end{itemize}

\begin{figure}[hbtp]
\centering
\includegraphics[scale=0.6]{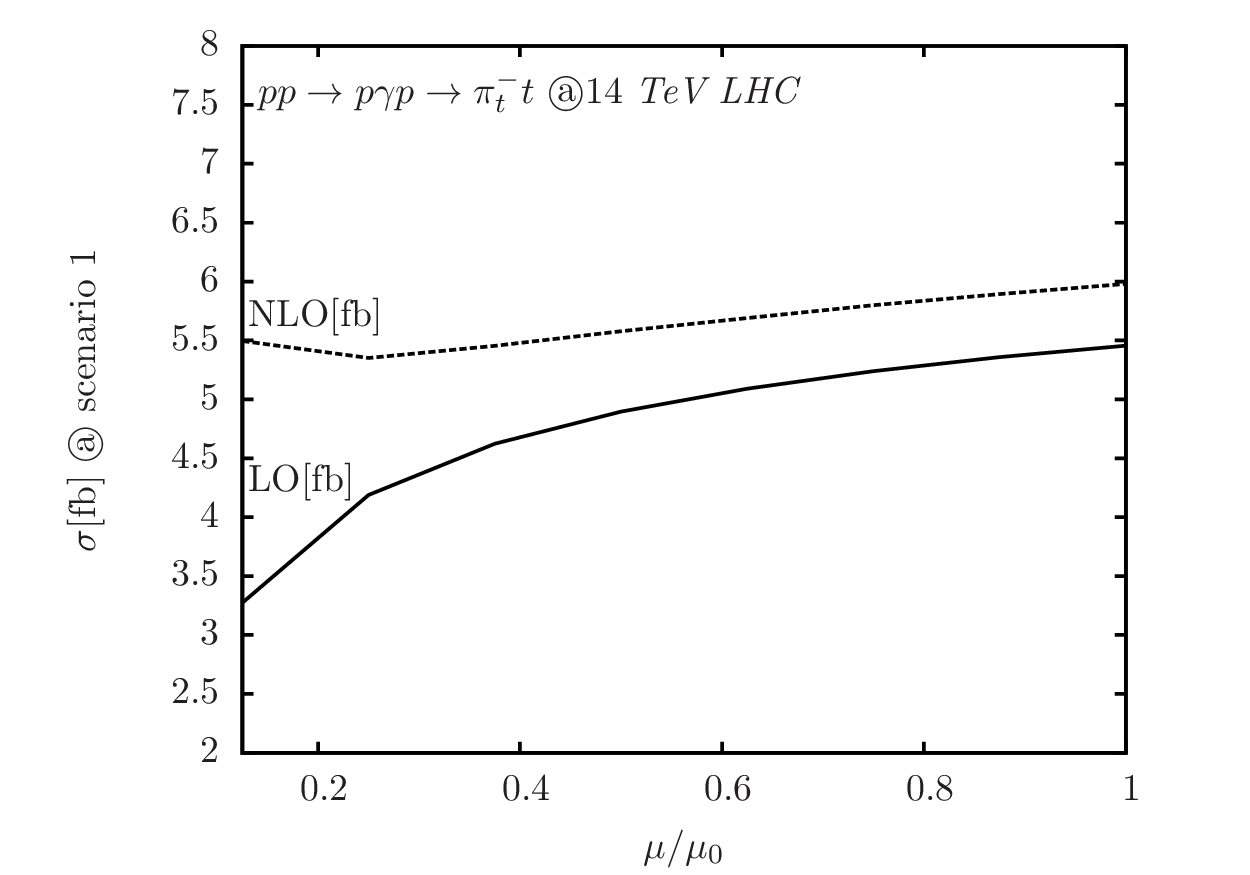}
\includegraphics[scale=0.6]{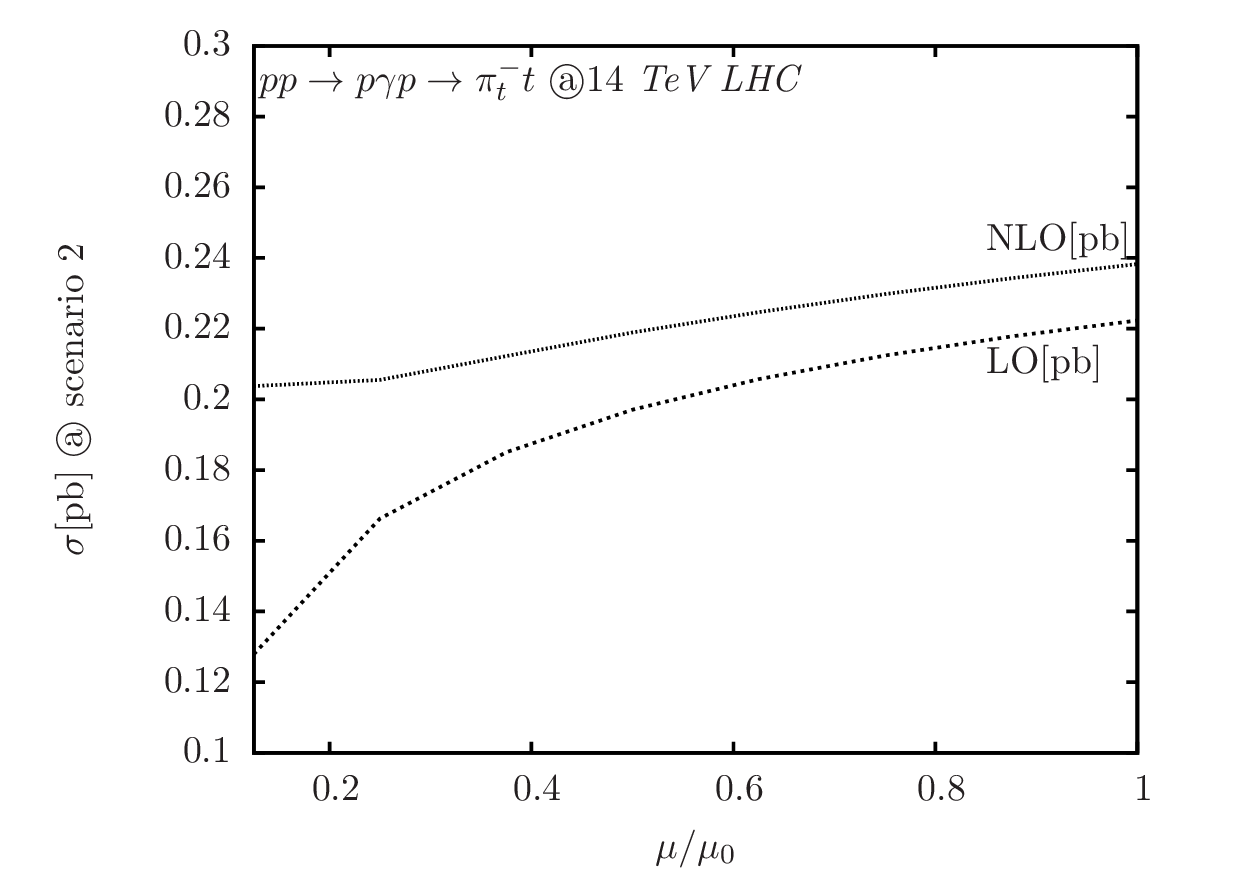}
\caption{\label{fig4}
The scale($\mu$) dependence of the $LO$ and $NLO$
$QCD$ corrected cross sections for $pp\rightarrow p\gamma p \rightarrow \pi^-_t t+X$
at the $\sqrt{s} = 14$ $TeV$ $LHC$ with $\mu_0=M_t$, $\delta_s=10^{-4}$ and $\delta_c=\delta_s/50$.
The experimental detector acceptances ($\xi_{min}<\xi<\xi_{max}$) are supposed to be
$0<\xi<1$. Solid and dashed lines for Scenario 1, $LO$ and $NLO$, respectively,
while dotted and dot-dotted lines for Scenario 2, $LO$ and $NLO$, respectively.
}
\end{figure}

Before presenting the numerical predictions, several checks should be done.
First, The $UV$ and $IR$ safeties are verified numerically after combining all
the contributions at the $QCD$ one loop level. We display random phase space points as
well as the cancelation for different divergent parameters
with the help of OneLoop\cite{Oneloop} to compare with our modified LoopTools.
Second, when do the phase space integration, we use
Kaleu\cite{Kaleu} to cross check especially for the hard emission contributions.
Third, since the total cross section is independent of the soft cutoff
$\delta_s(=\Delta E_g/E_b, E_b=\sqrt{\hat{s}}/2)$ and the collinear
cutoff $\delta_c$, trivial efforts should be done to check such independence.
Fourth, the scale ($\mu$) dependence should be reduced after considering the $NLO$ corrections.
Indeed, our results show that the scale uncertainty can be reduced obviously.
Choose the input scenario 1 as an example, if $\mu$ varies from $1/8\mu_0$ to $\mu_0=M_t$,
the $LO$ cross section varies from 3.2 $fb$ to 6 $fb$ while $NLO$ predictions stay much flat
between 5.5 $fb$ to 6.4 $fb$. For more details, see Fig.\ref{fig4}, where
we show the scale ($\mu$) dependence of the $LO$ and $NLO$ $QCD$ loop corrected cross sections for
$pp\rightarrow p\gamma p \rightarrow \pi^-_t t+X$.
In the further numerical calculations, we fix
$\delta_s = 10^{-4}$, $\delta_c=\delta_s/50$ and choose $\mu=\mu_0=M_t$.

\subsection{Cross sections and Distributions}

\begin{figure}[hbtp]
\centering
\includegraphics[scale=0.6]{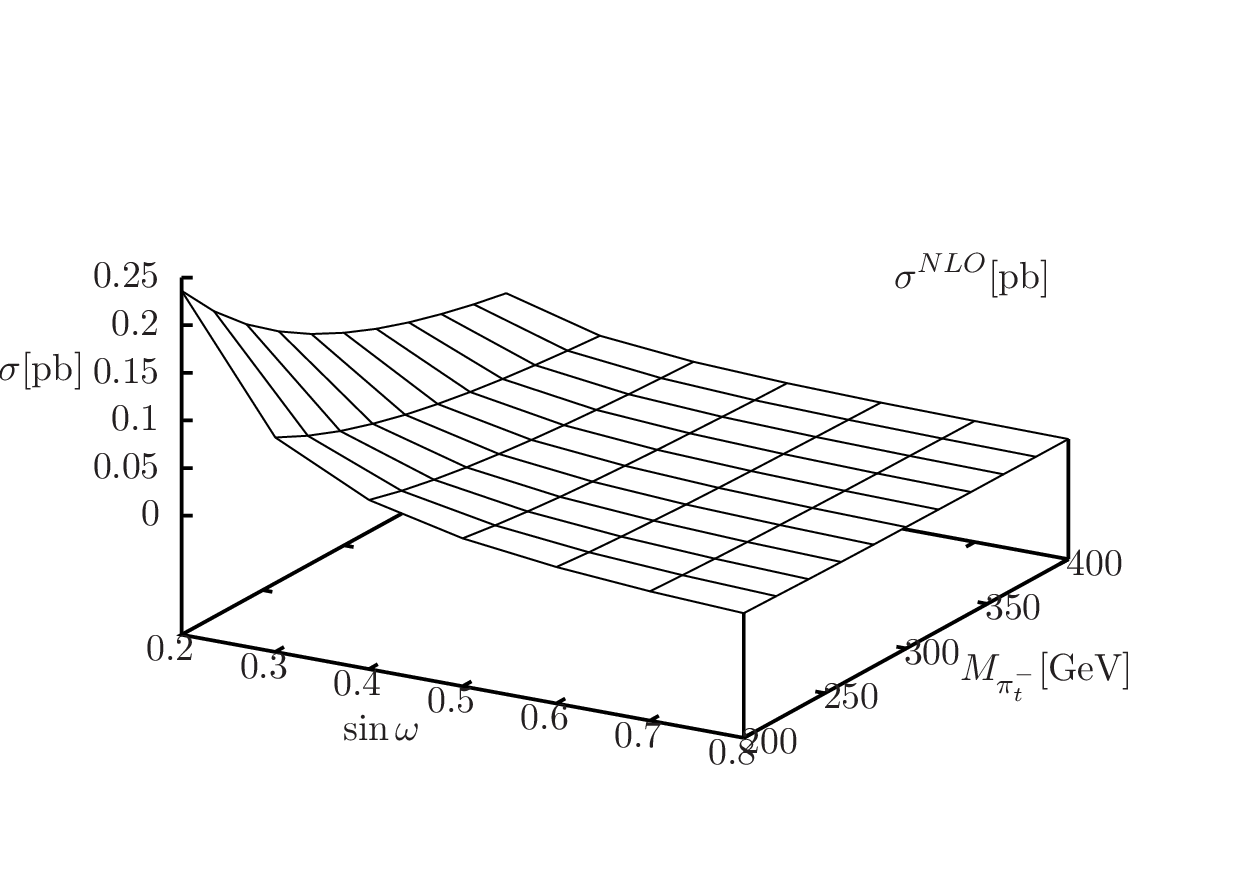}
\includegraphics[scale=0.6]{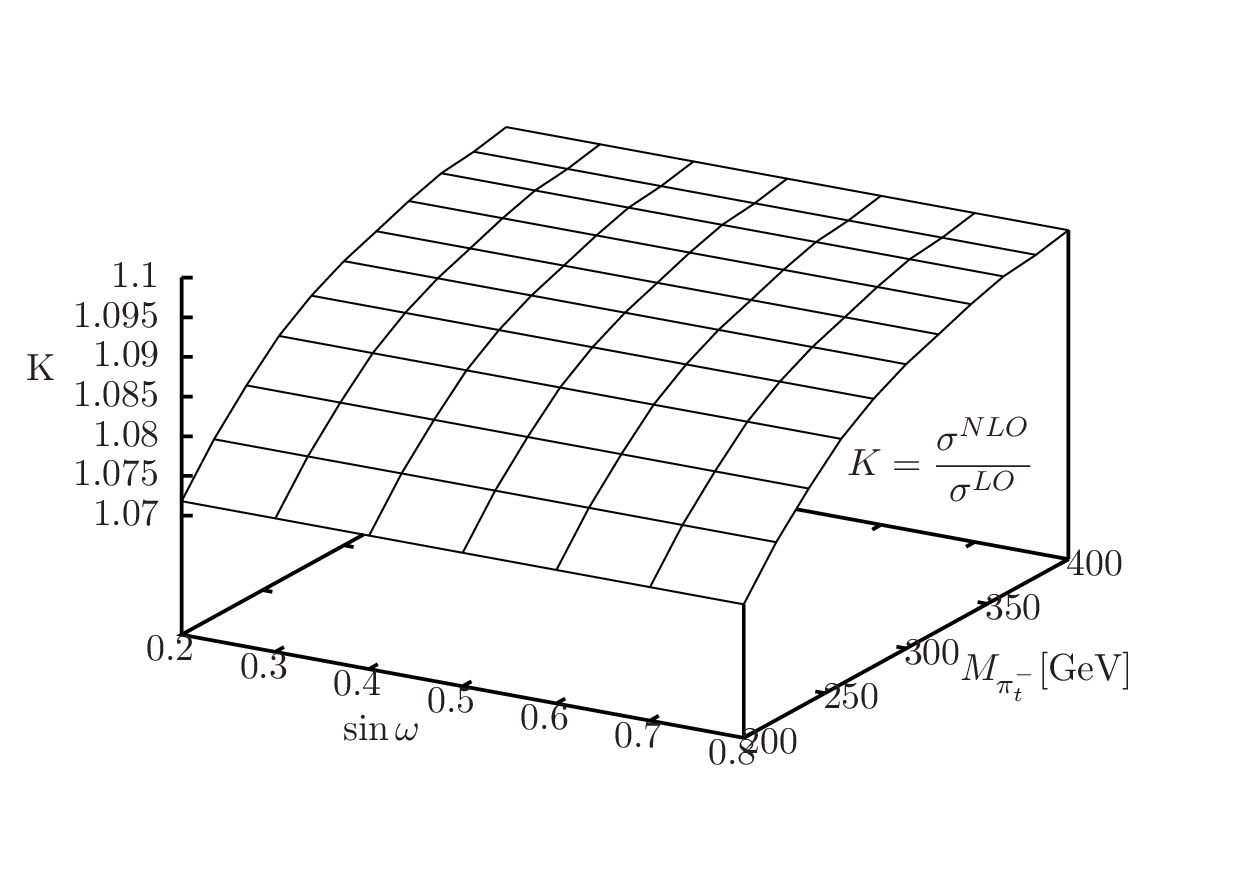}
\caption{\label{fig5}  Cross sections (the left panel) for $NLO$ predictions
and K-factor (the right panel) defined as $\sigma^{NLO}/\sigma^{LO}$
for $pp\rightarrow p\gamma p \rightarrow {\pi^-_t} t+X$
as functions of different values of parameters in $TTM$ models at 14 $TeV$ $LHC$.
Here we choose $0.0015<\xi<0.5$. The other parameters related to $TTM$ models
are chosen to be  $M_D=400\ GeV$, $M_{W'}=500$,
with $\sin\omega$ varies from 0.2 to 0.8 and $M_{\pi^-_t}$ from 200 to 400 $GeV$, respectively
, and the other $TTM$ model input parameters are chosen
to be scenario 2.}
\end{figure}

In Fig.\ref{fig5} we present the cross sections (the left panel) for $NLO$ predictions
and K-factor (the right panel) defined as $\sigma^{NLO}/\sigma^{LO}$
for $pp\rightarrow p\gamma p \rightarrow {\pi^-_t} t+X$
as functions of different values of input parameters in the $TTM$ model.
One is $\sin\omega$ and the other is the top-pion mass $M_{\pi_t}$.
Here we choose the detector acceptance as $0.0015<\xi<0.5$.
The other parameters related to the $TTM$ model
are chosen to be $M_D=400\ GeV$ and $M_{W'}=500$,
with $\sin\omega$ varies from 0.2 to 0.8 and $M_{\pi_t}$ from 200 to 400 $GeV$, respectively.
Our results show that the total $LO$ and $NLO$ cross sections are sensitive to
the input parameter $\sin\omega$. When $\sin\omega$ becomes larger, the cross sections reduce
obviously. Same behavior can be found for the charged top-pion mass $M_{\pi_t}$. When the mass
becomes heavier, the phase space of final states are suppressed thus leading lower cross sections.
The right panel presents the K-factor dependence on $\sin\omega$ and $M_{\pi_t}$. No matter how
$\sin\omega$ changes, the K-factor does not change much with a fixed top-pion mass.
While for $M_{\pi_t}$ become larger from 200 to 400 $GeV$, the K-factor grows up step-by-step, however,
not very much, see, from 1.07 to 1.1, leading the $NLO$ $QCD$ corrections up to around $7\% \sim 11\%$ within
our chosen parameters.

\begin{figure}[hbtp]
\centering
\includegraphics[scale=0.6]{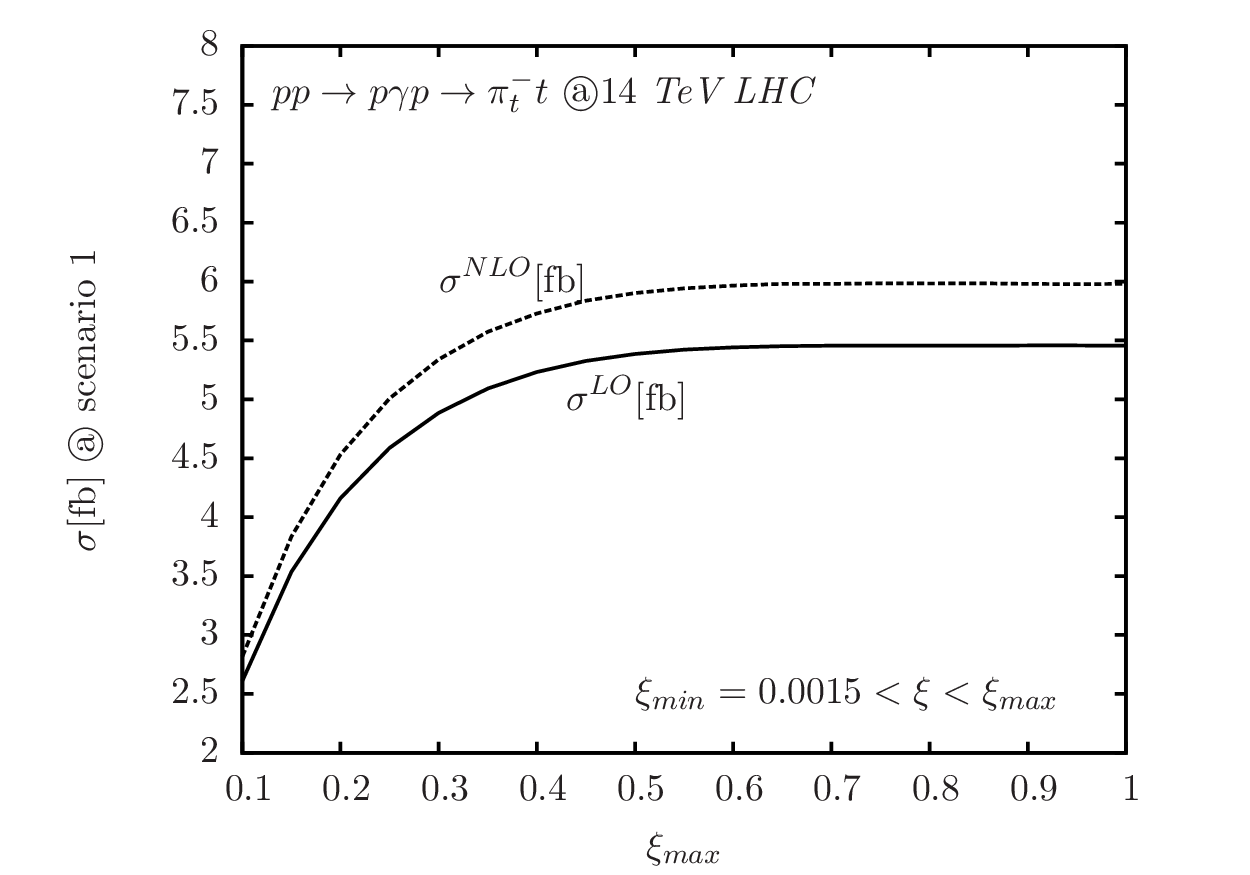}
\includegraphics[scale=0.6]{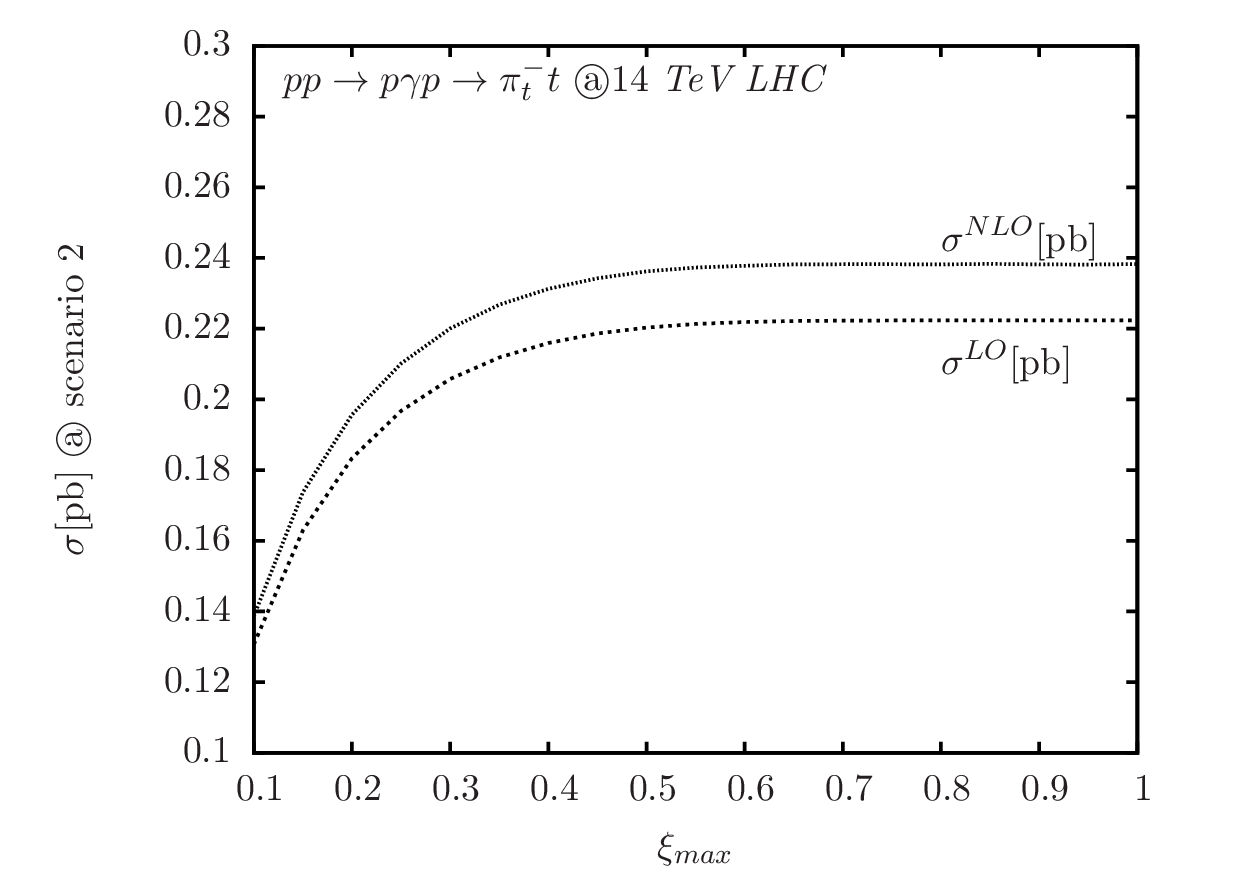}
\caption{\label{fig6}  Cross sections for $LO$ and $NLO$ predictions
for $pp\rightarrow p\gamma p \rightarrow \pi^-_t t+X$
as functions of different values of $\xi_{max}$ detector acceptances at the 14 $TeV$ $LHC$.
Here we fix $\xi_{min}=0.0015$ and take $\xi_{max}$ as a running parameter from 0.15 to 1.
Left panel with units in fb for $TTM$ scenario 1 with solid and dashed lines for $LO$ and $NLO$ predictions,
while right axis in pb for $TTM$ scenario 2 with dotted and dot-dotted lines for $LO$ and $NLO$, respectively.}
\end{figure}

To see how the cross sections depend on the detector acceptances,
in Fig.\ref{fig6} we fix $\xi_{min}=0.0015$ and take $\xi_{max}$ as a running parameter.
One should note here that the detector acceptance is indeed a step function of $\xi$
while here we show the dependence on $\xi$ qualitatively.
Cross sections for the two input scenarios
are presented as $\xi_{max}$ running from 0.15 to 1.
Left panel present results for scenario 1 with dotted and dot-dotted lines for $LO$ and $NLO$, while
right panel for scenario 2 with solid and dashed lines for $LO$ and $NLO$ predictions, respectively.
From these panels, we can see for $\xi_{max}<0.5$, the cross section enhance rapid when
$\xi$ acceptances become larger. Case is different for $\xi_{max}>0.5$ where little contributions contribute.
Furthermore, no matter how the detector acceptances changes,
the ratio of $\sigma^{NLO}$ to $\sigma^{LO}$ does not change much.
A typical value of K-factor equal 1.09 for scenario 1 and 1.07 for scenario 2,
lead the $NLO$ $QCD$ loop corrections up to $9\%$ and $7\%$ and keep unchange
as functions of running $\xi$.

\begin{figure}[hbtp]
\centering
\includegraphics[scale=0.6]{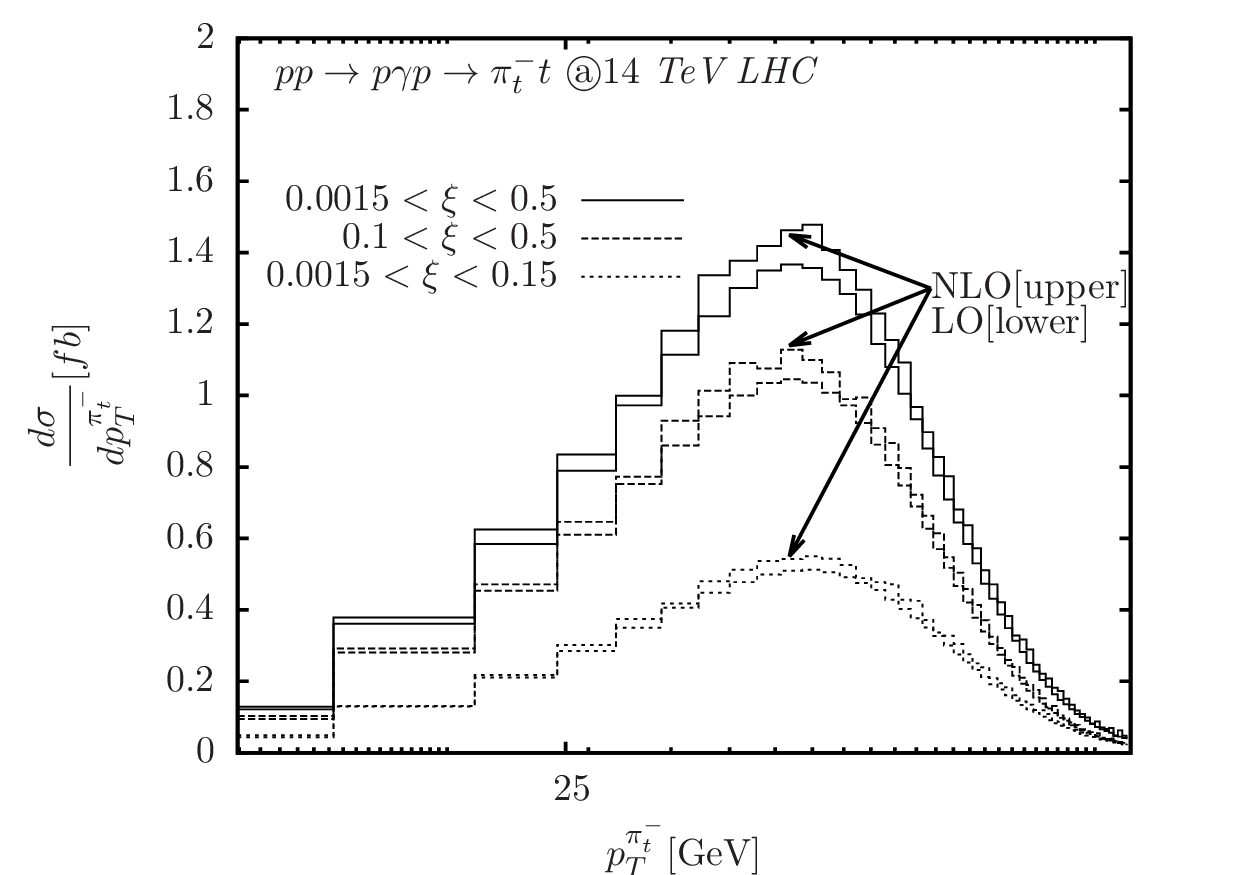}
\includegraphics[scale=0.6]{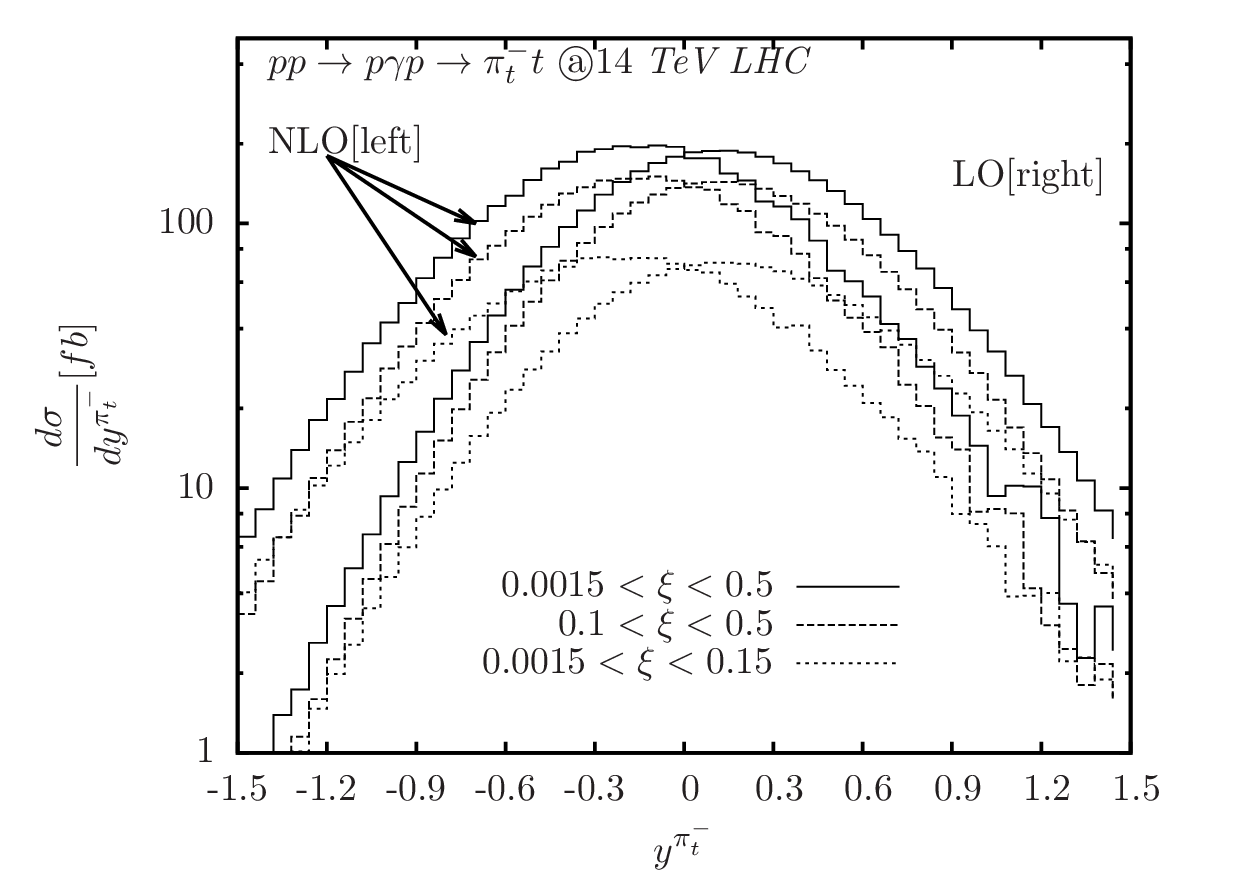}
\caption{\label{fig7}  The $LO$ (lower curves) and $NLO$ (upper curves) transverse momentum ($p_T$)
and Rapidity ($y$) distributions of the charged top-pion $\pi^-_t$ for the process
$pp\rightarrow p\gamma p \rightarrow \pi^-_t t+X$ at the 14 $TeV$ $LHC$. The experimental detector acceptances($\xi_{min}<\xi<\xi_{max}$) are chosen to be
$0.0015<\xi<0.5$ (solid lines), $0.1<\xi<0.5$ (dashed lines) and
$0.0015<\xi<0.15$ (dotted lines), respectively, and the $TTM$ model input parameters are chosen
to be scenario 2.}
\end{figure}

We present the transverse momentum ($p_T$) and rapidity ($y$) distributions
for the charged top-pion in Fig.\ref{fig7}.
For $p_T^{\pi^-_t}$, $NLO$ predictions can enhance
the $LO$ distributions obviously around the peak range and
the same behavior can be found for the $p_T^t$ distributions.
It will be interesting to see $y^{\pi^-_t}$
where the $NLO$ corrections can shift the $LO$ rapidity obviously
in the way of moving the position where $y^{\pi^-_t}$ peaked.
Take $0.0015<\xi<0.5$ as an example,
the distribution $y^{\pi^-_t}$ peaked at y=-0.18 for $LO$
while the $NLO$ predictions move the $LO$ $y^{\pi^-_t}$ peak to y=-0.42 but
no obvious enhancement to the $LO$ predictions.

\subsection{Signal Background Analysis and Parameter Sensitivity}
Now let's turn to the signal and background analysis.
From Ref.\cite{TTM1} we see that, for $M_{h_t} \geq 300GeV$ and $M_{\pi_t} \leq 600GeV$ ,
the charged top-pions $\pi^-_t$ dominantly decay into $\overline{t}b$ and
there is $Br(\pi^-_t \rightarrow \overline{t}b) > 90\%$.
As for the mass of $M_{\pi_t}$ become heavier, the validity of this statement
is no longer independent of the mass of, for example, top-Higgs mass $M_{h_t}$.
However, for each value of $\sin\omega$, a specific range of masses for the top-Higgs
is excluded by the Tevatron data. For example, the illustrative value $\sin\omega = 0.5$,
the data implies that the mass range $140 GeV<M_{H_t}<195 GeV$ is excluded.
Here we concentrate on the case where $M_{h_t} \geq 350GeV$.
Even though, as the mass $M_{\pi_t}$ become heavier than 600 $GeV$, the decay mode $\pi^\pm\rightarrow W^\pm H_t$
becomes more and more competitive, where the assumption of a branching ratio
$Br(\pi^-_t \rightarrow \overline{t}b) < 90\%$ should be considered.
We concentrate on the $\pi^\pm_t \rightarrow \overline{t}b(t\overline{b})$ decay modes.
In this case, photoproduction of the charged top-pion associated with a top quark
can easily transfer to the $t\bar{t}b$ final state through
\begin{eqnarray}
pp\rightarrow p\gamma p \rightarrow \pi^-_t t \rightarrow \bar{t} b  W^+ b  \rightarrow W^-\bar{b} b W^+ b \rightarrow
\ell^+\ell^- \bar{ b}b b  \slashed{E}_T
\end{eqnarray}
thus gives rise to the $\ell^+\ell^- \bar{ b}b b \slashed E_T$ signature via $\gamma b$ collisions at the  $LHC$.

The backgrounds appear in two kind of processes. The first, called irreducible background comes from photoproduction with very similar final state as the signal. The second has the same final state but occurs through different processes
induced by partonic interactions and is called reducible background. The key difference between photoprduction
and partonic interactions at the $LHC$ lies in the absence of colour exchange on the photon side.
This causes an important zone of rapidity to be completely devoid of hadronic activity called a large rapidity gap($LRG$) and which is natural way to distinguish photoproduction and partonic backgrounds.
In the framework of EPA, emitted quasi-real photons from the protons have a low virtuality and scattered with small angles from the beam pipe. Therefore when a proton emits a quasi-real photon it should also be scattered with a small angle. Hence, intact scattered protons exit the central detector without being detected. This causes a decrease in the energy deposit in the corresponding forward region compared to the case in which the proton remnants are detected by the calorimeters. Consequently, for any reaction like $pp\rightarrow p\gamma p\rightarrow pX$, one of the forward regions of the central detector has a significant lack of energy. The region with a lack of  energy (or equivalently lack of particles) defines a forward $LRG$. Backgrounds from usual $pp$ deep inelastic processes can be rejected by applying a selection cut on this quantity.

In addition, another tagging method based on the same physics properties of photoproduction events is to place an exclusivity condition on reconstructed particle tracks on the gap side which can obviously reduce patronic backgrounds\cite{singleTphoto}. Even if both conditions are used and partronic background is reduced to a level that not allows proper signal extraction, elastic photon emission can be tagged using very forward detector ($VFD$)\cite{VFD} placed hundreds of meters away from the interaction point. For instance, the case for which $VFD$ stations would be put at 220m and 420m from the interaction point and is mandatory in order to retain partonic backgrounds low\cite{WHphotoproduction}. Indeed, when an intact proton is scattered with a large pseudorapidity it escapes detection from the central detectors. But since its energy is lower than the beam energy, its trajectory decouples from the beam path into the very forward region. Forward detectors can detect particles with a large pseudorapidity. The detection of final state intact protons by the forward detectors provides a characteristic signature. Backgrounds from usual $DIS$
processes can also be rejected by use of this characteristic signature provided by the forward detectors.

Therefore in our paper, the only considered backgrounds come from protoproduction.
From this point we can see that the backgrounds would come from $t\bar{t}$ plus $jet$ ($t\bar{t}j$) photoproduction.
Different from normal $pp$ collision, in $\gamma p$ collisions where photoproduction of top quark pairs has
similar cross sections like, for example, $W^- t$ productions, only $\sim$1.4 pb\cite{HEPhotonIntatLHC},
while for $t\bar{t}j$, roughly $\sim$ 16fb after considering the fake b-tagging efficiency,
leading such related background processes easier going than in case of the $pp$ collision.
Here we assume that the $\pi^\pm_t$ fully decay to $t\bar{b}(\bar{t}b)$ if
$M_{\pi_t} < 600GeV$ while $Br(\pi^\pm_t \rightarrow t\bar{b}(\bar{t}b) < 90\%)$ \cite{TTM1} should be considered if
$M_{\pi_t} \geq 600GeV$. For the $SM$ gauge bosons $W^{\pm}$  decay leptonically, $W^{\pm}\rightarrow l\nu$, the signal is $S={\cal L} \times \sigma(pp\rightarrow prp \rightarrow \pi^\pm_t t\rightarrow t\bar{t}b)\times
K^{NLO} \times [BR(t\rightarrow Wb)]^2
\times [3\times BR(W\rightarrow \ell \nu)]^2$ and the corresponding background as
$B={\cal L} \times \sigma(pp\rightarrow prp \rightarrow t\bar{t}j)\times Eff_{j} \times [BR(t\rightarrow Wb)]^2
\times [3\times BR(W\rightarrow \ell \nu)]^2$ with $j=u,d,c,s, b, \bar{u},\bar{d},\bar{c},\bar{s},\bar{b},g$
and $Eff_{j}$ is the fake b-tagging efficiency of the jets.
For c-jets and light jets, a fake b-tagging efficiency of 10$\%$ and 1$\%$ respectively is assumed.
Here we take $BR(t\rightarrow Wb) \approx 1$ and $BR(W\rightarrow \ell \nu) \approx 0.108$.
For the luminosity ${\cal L}$ we take $1 fb^{-1}$, $10 fb^{-1}$, $100 fb^{-1}$, respectively.
In Table.\ref{tab3}, we present the parameters sensitivity on the signal background ratio $S/\sqrt{B}$.
Here we choose $0.0015<\xi<0.5$. The background cross section after consider all the
b-tagging efficiency and the rejection factors for the $c, \bar{c}$ and light jets is 1.68 $fb^{-1}$.
The $5\sigma$ and $3\sigma$ bounds of the parameters are presented with three values of the luminosity.
For $S/\sqrt{B}>5$, the new physics signal will be detected obviously while for $S/\sqrt{B}<3$
it will be challenge to be detected.

Our results show that, for low $M_{\pi_t}$,
the $\sin\omega$ discovery range is larger than the case of high $M_{\pi_t}$.
As the top-pion mass becomes larger, the $\sin\omega$ discovery range is suppressed.
When $M_{\pi_t} > 900 GeV$, heavy final state strongly suppress the phase space. The
signal becomes much small and makes it more challenge to be detected. In this case,
higher luminosity is needed to make the detection possible and push the discovery boundary larger.
Two ways can be used in order to constraint the parameters or the excluding boundary more strictly:
one is, as we see, to enhance the luminosity which can expand the related parameter space, see in Table.\ref{tab3},
while the other one is to take more kinematical cuts to improve the ratio $S/\sqrt{B}$.
In our case for example, if a $p_T^{jet}$ cut taken to be larger than 200 $GeV$ can strongly suppress the $ttj$ backgrounds and thus lead better $S/\sqrt{B}$ in parts of the $TTM$ parameter space.

\begin{table}
\begin{center}
\begin{tabular}{l c c c c c c c c c}
\hline\hline
  && \multicolumn{8}{c}{$\sin\omega$} \\
$M_{\pi_t}$  & &\multicolumn{2}{c}{${\cal L} = 1 fb^{-1}$}&&\multicolumn{2}{c}{${\cal L} = 10 fb^{-1}$}&&\multicolumn{2}{c}{${\cal L} = 100 fb^{-1}$}\\
$[GeV]$  &  &  $5\sigma$     &    $3\sigma$    && $5\sigma$     &    $3\sigma$ && $5\sigma$     &    $3\sigma$\\
\hline
300   &&  0.596   &  0.694  &&  0.800  &  0.867  &&  0.923   &  0.950   \\
400   &&  0.450   &  0.546  &&  0.671  &  0.762  &&  0.851   &  0.902   \\
500   &&  0.340   &  0.423  &&  0.543  &  0.642  &&  0.758   &  0.833   \\
600   &&  0.258   &  0.327  &&  0.431  &  0.526  &&  0.650   &  0.743   \\
700   &&  0.183   &  0.231  &&  0.308  &  0.386  &&  0.500   &  0.599   \\
800   &&  0.137   &  0.172  &&  0.232  &  0.292  &&  0.388   &  0.478   \\
900   &&  $<0.1$   &  0.131  &&  0.174  &  0.220  &&  0.294   &  0.370   \\
1000   &&  $<0.1$   &  $<0.1$  &&  0.132  &  0.164  &&  0.223   &  0.282   \\
1100   &&  $<0.1$   &  $<0.1$  &&  $<0.1$  &  0.123  &&  0.165   &  0.208   \\
1200   &&  $<0.1$   &  $<0.1$  &&  $<0.1$  &  $<0.1$  &&  0.122   &  0.149   \\
\hline\hline
\end{tabular}
\end{center}
\vspace*{-0.8cm}
\begin{center}
\begin{minipage}{14cm}
\caption{\label{tab3} The $TTM$ parameters $\sin\omega$ and $M_{\pi_t}$ sensitivities on the signal background ratio $S/\sqrt{B}$.
$5\sigma$ for the discovery boundary and $3\sigma$ for the excluding boundary. The detector acceptance here is
chosen to be $0.0015<\xi<0.5$.}
\end{minipage}
\end{center}
\end{table}

\vskip 5mm
\section{Summary}
\par
In this work, we present the precise photoproduced charged top-pion $\pi^\pm_t$ production
associated with a top through $pp\rightarrow p\gamma p \rightarrow \pi^\pm_t t+X$
at the 14 $TeV$ $LHC$ at NLO QCD loop level.
We find the cross sections are sensitive to $TTM$ parameters,
and the smaller the $\sin\omega$ is or the lighter the top-pion $\pi^-_t$ is,
the larger the cross sections will be.
The typical $QCD$ correction value is $7\% \sim 11\%$
which does not depend much on the $TTM$ parameter $\sin\omega$ as well as
the detector acceptances $\xi$.
We also present the $5\sigma$ discovery and $3\sigma$ excluding boundaries
as functions of the $TTM$ parameters for three values of the luminosity at the future $LHC$.

\section*{Acknowledgments} \hspace{5mm}
Sun Hao thanks Dr. Inanc Sahin for his kindness to provide invaluable advice.
Project supported by the National Natural Science Foundation of China
(No. 11205070, 11275088), Shandong Province Natural Science Foundation (No. ZR2012AQ017),
Natural Science Foundation of the Liaoning Scientific Committee (No. 201102114),
Foundation of Liaoning Educational Committee (No. LT2011015)
and by the Fundamental Research Funds for the Central Universities (No. DUT13RC(3)30).

\end{document}